\documentclass[12pt]{article}

\usepackage{scicite}
\usepackage{graphicx}
\usepackage{array}
\usepackage{amsmath,amsfonts,amssymb}
\usepackage[ansinew]{inputenc}
\usepackage{color}
\usepackage{calc}
\usepackage{times}

\newenvironment{sciabstract}{%
\begin{quote} \bf}
{\end{quote}}

\title{Electrofluorochromism at the single molecule level}

\author
{Benjamin Doppagne$^1$, Michael C. Chong$^1$,Herv\'e Bulou$^1$, Alex Boeglin$^1$,\\
Fabrice Scheurer$^1$, Guillaume Schull$^{1\ast}$\\
\normalsize{$^1$ Universit\'e de Strasbourg, CNRS, IPCMS, UMR 7504, F-67000 Strasbourg, France,} \\
\\
\normalsize{$^\ast$To whom correspondence should be addressed; E-mail:  schull@unistra.fr}
}

\date{}

\begin{document}

\baselineskip24pt

\maketitle

\begin{sciabstract}
The interplay between the oxidation state and the optical properties of molecules plays a key role for applications in displays, sensors or molecular-based memories. The fundamental mechanisms occurring at the level of a single-molecule have been difficult to probe. We used a scanning tunneling microscope (STM) to characterize and control the fluorescence of a single Zn-phthalocyanine radical cation adsorbed on a NaCl-covered Au(111) sample. The neutral and oxidized states of the molecule were identified on the basis of their fluorescence spectra that revealed very different emission energies and vibronic fingerprints. The emission of the charged molecule was controlled by tuning the thickness of the insulator and the plasmons localized at the apex of the STM tip. In addition, sub-nanometric variations of the tip position were used to investigate the charging and electroluminescence mechanisms.
\end{sciabstract}

\section*{}

The fluorescence of electrofluorochromic molecules depends on their charge state \cite{Audebert2013,Beneduci2014 } and is responsive to electrical stimuli. The optical properties of oxidized or reduced molecular species are generally addressed by spectroelectrochemistry \cite{Kaim2009} for large ensembles of molecules within an electrochemical cell. This allows optical characterization while switching the oxidation state of a large number of molecules. Recent progress has shown that single-molecule sensitivity can be reached with this approach \cite{Hill2013}, but vibronic spectroscopy from molecular fluorescence \cite{Lei2008,Weichun2017} has not been reported. At the single molecule limit and in an electrochemical environment, such vibronic signals were only observed in surface-enhanced Raman spectroscopy measurements \cite{Cortes2010,Zaleski2015}, but at the cost of a direct contact between the molecule and the electrode that may alter the molecular properties. More importantly, it is not possible to obtain direct information regarding the immediate environment of the molecule, nor reach sub-molecular optical resolution with any of these approaches.

Scanning probe microscopies can be used to control the charge state of atoms and molecules \cite{Repp2004,Wu2006,Ingmar2011,Leoni2011,Torrente2012,Fatayer2018} and excite their vibrationaly resolved fluorescence \cite{Qiu2003,Chen2010, Lee2014, Chong2016, Chong2016a, Doppagne2017, Imada2017} with sub-molecular precision. Here we combine these approaches and report on the STM-induced fluorescence of a ZnPc molecule in its neutral and (radical) cationic states. The luminescence spectra, obtained for ZnPc molecules decoupled from a Au(111) sample by thin layers of NaCl, reveal the electronic and vibronic signatures of the two oxidation states with a high spectral resolution. These data suggest a fast blinking of the molecule between its neutral and cationic form. Varying the exact number of insulating layers and the plasmonic response of the tip-sample cavity allowed us to manipulate the charge and excited states lifetimes. Finally, subnanometric variations of the tip position with respect to the molecule provided additional clues regarding the charging and fluorescence excitation mechanisms of the molecule. 
\noindent

\noindent 
\begin{figure*}
\includegraphics[width=151mm]{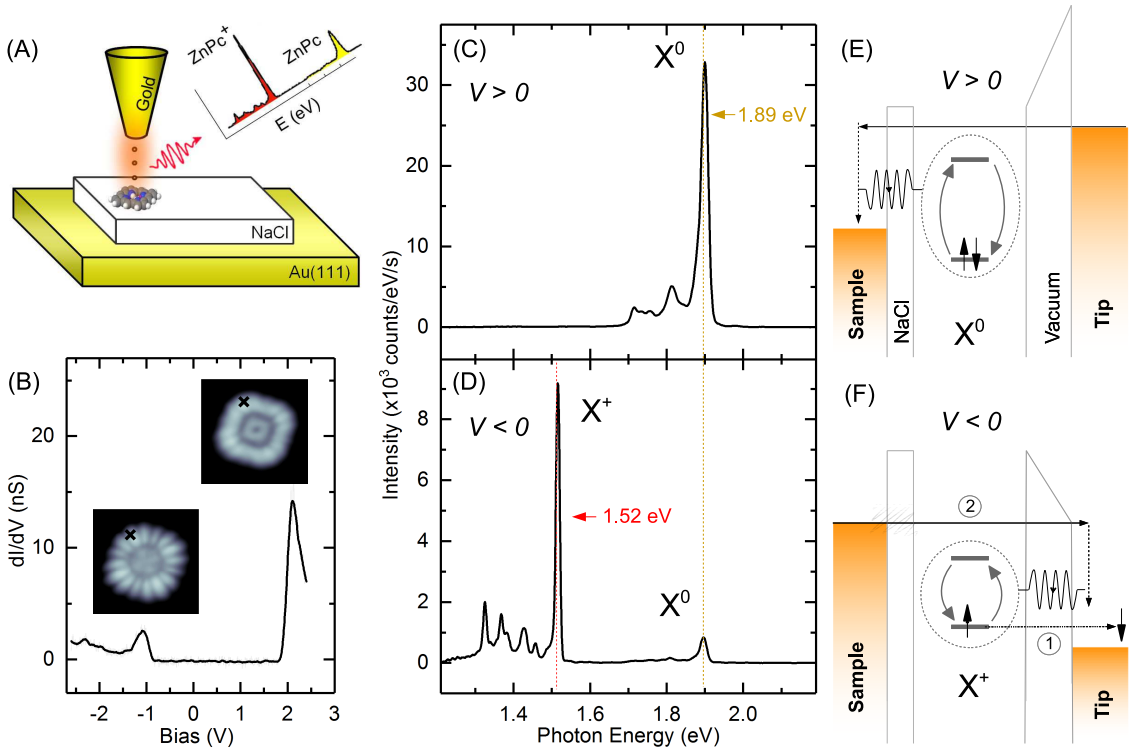}
\caption{\label{fig1}\textbf{STM-induced fluorescence of neutral and charged ZnPc molecules.} (A) Sketch of the STM-induced emission experiment. (B) d$I$/d$V$ spectrum acquired on a single ZnPc adsorbed on a trilayer of NaCl on Au(111). STM images (inset, 2.9$\times$2.9 nm$^{2}$, $I$= 30 pA) acquired at $V$ = -1 V (left) and $V$ = 1.85 V (right). (C and D) STM-LE spectra acquired at positive voltage (C, $V$ = 2 V, $I$ = 300 pA, acquisition time $t$ = 300 s) and negative voltage (D, $V$ = -2.5 V, $I$ = 300 pA, $t$ = 300 s) for the STM tip located at the positions marked by black crosses in the images in (B). Sketches of the luminescence mechanisms for the neutral (E) and charged (F) ZnPc. The molecular states reflect the optical gap of the neutral and charged molecule.}
\end{figure*}
\noindent
\\

The electronic and electroluminescent (EL) properties of single ZnPc molecules separated from a Au(111) surface by three layers of NaCl are shown in Fig.1(A). The d$I$/d$V$ spectrum acquired on a single molecule (Fig.1(B)) revealed an electronic gap of 3.2 eV between the highest occupied molecular orbital (HOMO) and the lowest unoccupied molecular orbital (LUMO). These orbitals can be readily identified in STM images (inset) acquired at the corresponding energies. The same energy gap was reported for ZnPc deposited on salt-covered Ag(111) samples \cite{Zhang2016,Doppagne2017}, although with a rigid shift of the molecular orbitals to higher energies ($\approx$ 1 eV) reflecting the higher work function of Au(111) ($\approx$ + 0.8 eV) with respect to Ag(111). STM-induced light emission spectra acquired at positive (+ 2.5 V, Fig.1(C)) and negative (- 2.5 V, Fig.1(D)) sample voltages revealed a sharp ($\approx$ 20 meV) emission line at 1.89 eV (labelled X$^{0}$) characteristic of the fluorescence of a single ZnPc molecule in its neutral form \cite{Zhang2016,Doppagne2017,Zhang2017}. The emission is 30 times more intense at positive voltage where low energy vibronic features are also observable. Similar light emission features \cite{Schneider2012a,Chong2016,Imada2016,Doppagne2017,Zhang2017,Imada2017} were assigned to an excitation of a decoupled molecule by an energy transfer from an inelastic tunneling electron, possibly mediated by localized plasmons (see sketch Fig.1(E)).

The spectrum in Fig.1(D) also displayed an intense and sharper ($\approx$ 10 meV) emission line at 1.52 eV (labelled X$^{+}$), that was not reported in previous STM-induced luminescence measurements. This value is in excellent agreement with the fluorescence of ZnPc radical cations (ZnPc$^{+}$) reported by experimental and theoretical studies \cite{Nyokong1987,Mack2006,Rosa2009}. Fig.1(F) presents a possible explanation for this observation. For a sufficiently high negative voltage, the Fermi level of the tip becomes resonant with the singly occupied molecular orbital (SOMO) of the ZnPc allowing for the tunneling of an electron from the molecule to the tip [(1) in Fig.1(F)]. Because the molecule is decoupled from the metallic substrate by a thin insulating layer, its radical cation may be stabilized long enough for other tunneling events [(2) in Fig1(F)] to occur. The energy lost by one of these electrons may then be transferred to the charged molecule that undergoes an excitation/emission cycle. The radical cation can only be generated at negative voltage which explains why the X$^{+}$ line is not observed at positive voltage \cite{note1}. The red-shifted fluorescence of ZnPc$^{+}$ with respect to ZnPc then follows from the intrinsic changes in the electronic structure upon removal of one electron from the macrocycle \cite{Rosa2009}. The widths of the emission lines are not lifetime limited \cite{note3} but rather reflect the dephasing caused by the interactions of the emitter with surface phonons. Indeed, density functional theory (DFT) calculations performed on an extremely close system \cite{Fatayer2018}, revealed that charging the molecule led to an important reorganization of the NaCl below the molecule. This hints at a reduction of the dephasing due to the molecule-salt couplings upon charging. The proposed luminescence mechanism is discussed further in the light of the current and voltage dependencies of the X$^{0}$ and X$^{+}$ contributions in S1 and S2 of the Supplementary Materials.

The observation of charged atoms and molecules was reported for several double tunneling junctions \cite{Repp2004,Wu2006,Ingmar2011,Leoni2011,Torrente2012,Fatayer2018}. It is generally associated to a sharp feature in the d$I$/d$V$ spectra at voltages characteristic of the charging energy. The absence of such a feature in the conductance spectrum Fig.1(B)) together with the simultaneous observation of X$^{0}$ and X$^{+}$ in the optical spectrum Fig.1(D) suggests a fast blinking between the neutral and the charged states. The frequency of the blinking is $>$ 1 kHz, the limit of our setup. Note that the existence of a charged state in our experiment can only be deduced from the optical measurements and would be missed on the basis of the STM/STS experiments.

\begin{figure}
\includegraphics[width=0.6\linewidth]{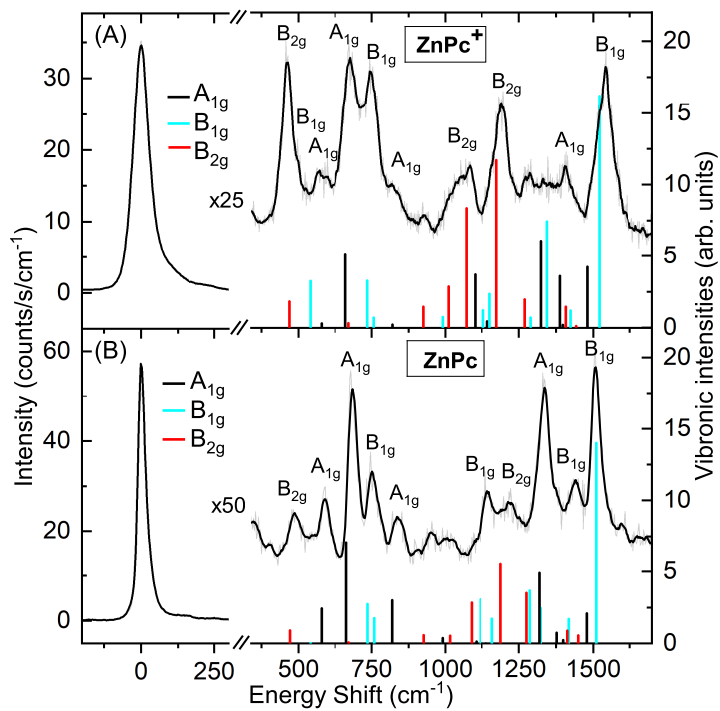}
\caption{\label{fig2}\textbf{Vibronic fingerprints of neutral and charged ZnPc molecules.} STM-LE spectra (black lines) revealing the vibronic spectroscopic fingerprint of (A) a single ZnPc radical cation deposited on a trilayer of NaCl on Au(111) ($V$ = -2.5 V, $I$ = 300 pA, $t$ = 180 s) and (B) of a neutral ZnPc (adapted with permission from \textit{(\citenum{Doppagne2017}) (copyright by the American Physical Society.)}) belonging to a molecular tetramer deposited on a trilayer of NaCl on Ag(111) \cite{note2} ($V$ = -2.5 V, $I$ = 750 pA, $t$ = 300 s). The colored bars correspond to theoretical vibronic intensities for the A$_{1g}$ (black), B$_{1g}$ (cyan), B$_{2g}$ (red).} 
\end{figure} 
\noindent

The spectrum in Fig.1(D) also revealed several vibronic features on the low energy side of the X$^{+}$ peak. An enlarged view of this spectrum (Fig.2(A)) represents the vibronic lines as a function of their shift in energy with respect to the X$^{+}$ line. A comparison with the vibronic spectrum characteristic of the neutral molecule (Fig.2(B) \cite{note2}) revealed that the spectroscopic fingerprint of the ZnPc molecule is strongly affected by its redox state. We found that vibronic lines were only slightly shifted in energy compared to the neutral case, but their relative intensities were strongly modified. DFT geometry optimizations of ZnPc showed changes of less than 0.01 {\AA} in bond length upon oxidation (see S3 in Supplementary Materials), highlighting the large rigidity of the ZnPc backbone and explaining the absence of shift of the vibronic lines. To understand the changes in peak intensities, we have developed a finite difference method based on time-dependent density functional theory calculations (see Materials and Methods in Supplementary Materials for details). We calculated the Frank and Condon factors of A$_{1g}$ vibrational modes as well as the Herzberg-Teller contribution to the transition moment of the first excited states for A$_{1g}$, B$_{1g}$ and B$_{2g}$ modes and hence evaluated the respective intensities of each vibronic mode for both oxidation states. For the charged molecule the B$_{2g}$ lines are enhanced and the A$_{1g}$ lines attenuated. In contrast to inelastic electron tunneling spectroscopy where the selection rules leading to the presence of vibrational features in tunneling spectra are generally poorly defined \cite{reed2008}, our fluorescence spectra provide a robust vibronic fingerprint of the probed molecule and its redox state. Similar vibronic information averaged over a large number of charged molecules has previously been obtained from Raman spectroscopy, but this is the first example where such a spectroscopic fingerprint is observed in the fluorescence spectrum of an isolated molecule.  
\\
\begin{figure}
\includegraphics[width=0.6\linewidth]{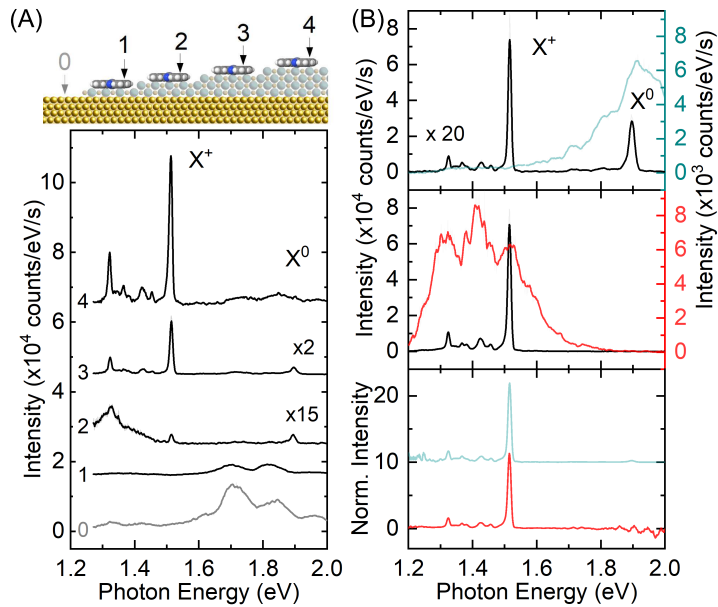}
\caption{\label{fig3} \textbf{Tuning the luminescence intensities.} (A) STM-LE spectra acquired, with the same STM tip, on the bare Au(111) surface ((0), $V$ = -2.5 V, $I$ = 300 pA, $t$ = 60 s ) and on single ZnPc molecules ($V$ = -2.5 V, $I$ = 300 pA) separated from the Au(111) surface by an increasing number of NaCl layers (1ML ($t$ = 300 s), 2ML ($t$ = 180 s), 3ML ($t$ = 120 s) and 4ML ($t$ = 10 s) see sketch). The spectra labelled 0 to 4 have been vertically shifted for clarity. (B) Two STM-LE spectra (black lines, $V$ = -2.5V, $I$= 300 pA, $t$ = 300 s (top panel), $t$ = 60 s (middle panel)) acquired on single ZnPc on 3ML of NaCl on Au(111) with two different STM tips, whose plasmonic response is deduced from STM-LE spectra acquired in front of the NaCl/Au(111) surface (red lines, $V$ = -2.5 V, $I$ = 300 pA, $t$ = 120 s; green lines, $V$ = -2.5 V, $I$ = 300 pA, $t$ = 60 s). The bottom panel displays the molecular emission spectra normalized by the respective plasmonic responses of the junction.}
\end{figure} 
\noindent

We followed the evolution of the EL spectrum recorded at a negative voltage on a single ZnPc molecule as a function of the number of NaCl atomic layers separating the molecule from the Au(111) surface. For a molecule adsorbed on a single NaCl layer, the STM-LE spectrum ((1) in Fig.3(A)) was similar to the one characteristic of the localized plasmon directly recorded on Au(111) ((0) in Fig.3(A)), indicating a quenched molecular emission. The X$^{0}$ and X$^{+}$ emission lines appeared in the spectra acquired on molecules separated from the metal by at least two atomic layers of salt ((2) in Fig.3(A)). The two contributions had similar intensities, but the emission of the radical cation strongly dominates for three ((3) in Fig.3(A)) and four ((4) in Fig.3(A)) salt layers. We believe that the relative intensity of the X$^{0}$ and the X$^{+}$ reflects the time spent by the molecule in the neutral and charged states. Indeed, increasing the molecule-sample distance reduced the neutralization probability of the ZnPc radical cation, while the tip-molecule distance that is slightly reduced to compensate for the larger salt barrier, increased the charging probability. Thus, the molecule spent more time in its cationic state when the number of salt layer increased \cite{note4}. 

Another striking observation was that, excepted for the X$^{0}$ peak that vanished in the 4ML spectrum, the absolute emission intensity of the two spectral contributions increased as a function of the number of NaCl layers. Overall, this change reflected the progressive reduction of the quenching of the molecular excitons caused by the hybridization with the sample states as the number of NaCl layer increased, and the large field enhancement \cite{Marinica2012} expected for plasmonic gaps of $\approx$ 1 nm. The disappearance of the X$^0$ peak in the 4ML spectrum suggests that, for this large molecule-sample distance, the molecule spent most of the time in its cationic form. We could also tune the relative intensity of the X$^{0}$ and the X$^{+}$ lines by changing the spectral response of the plasmons localized at the tip-sample junction (Fig.3(B)). As expected \cite{Qiu2003,Chong2016}, the presence of a plasmon resonance at the energy of the molecular emission lines amplifies their emission, a phenomenon explained by a stronger Purcell effect. Normalizing the molecular spectra by those of the plasmons (bottom panel of Fig.3(B)) revealed a linear dependence of the emission line intensities with plasmon intensity, confirming the above interpretation. 
\\ 
\begin{figure*}
\includegraphics[width=151mm,clip=]{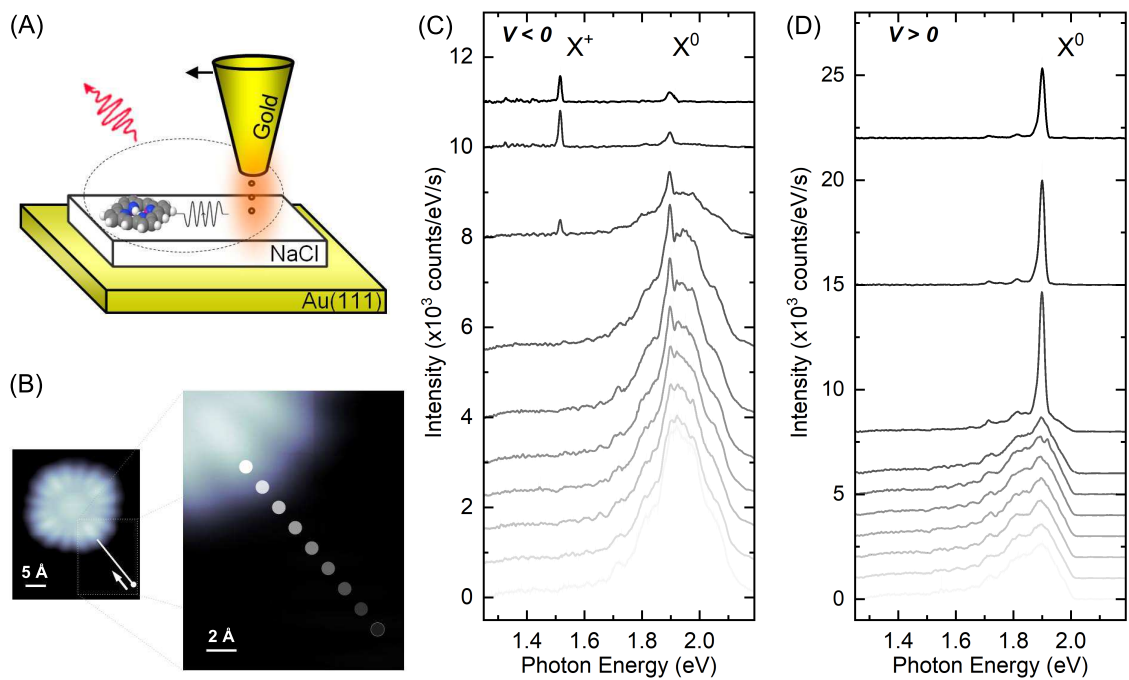}
\caption{\label{fig4} \textbf{Distance-dependent measurements of the luminescence contributions.} STM-LE spectra acquired for subnanometric variations of the lateral position of the STM tip with respect to a single ZnPc molecule (sketched in (A)) for negative voltage ((C), $V$ = -2.5 V, $I$ = 180 pA , $t$ = 180 s) and for positive voltage ((D), $V$ = 2 V, $I$ = 60 pA , $t$ = 180 s). The position of the tip for each spectrum is marked by a colored dot in the STM image ($I$ = 30 pA, $V$ = -2.5 V) in (B).}
\end{figure*} 
We recorded the evolution of the optical spectrum as a function of subnanometric variations in the lateral distance separating the STM tip from a ZnPc molecule (fig.4(A, and B)). At negative voltage (Fig.4(C)), a Fano-like feature was observed at the energy of the X$^{0}$ contribution even when the tip is located on top of the NaCl layer at $\approx$ 0.5 nm from the molecule edge. This response, first reported on in \cite{Imada2016} and later in \cite{Zhang2017,Imada2017}, reflects the coupling between the localized plasmons and the molecular exciton. Except for a subtle modification of the peak shape, this contribution did not change when the tip was moved on top of the molecule, suggesting that the fluorescence mechanism remained the same in this case. The response was rather different for the X$^{+}$ line, which only appeared when the tip was located on top of the molecule (Fig.4(B)). Indeed, while the plasmon/exciton coupling may have a measurable impact on the spectrum even when the tip is not located directly on top of the molecule, the oxidation reaction can only take place when an electron tunnels from the molecule to the tip, which requires a direct overlap between the tip and molecular orbitals. Finally, we performed the same experiment at positive voltage (Fig.4(D)). In this case, surprisingly, the X$^{0}$ contribution was only observed when the tip and molecular orbitals overlapped, in contrast with the results observed at negative voltage. The voltage dependency of this contribution revealed a multiple electron excitation process (see S4 in Supplementary Materials), an observation that hints at an excitation of the molecule by charge rather than energy transfer at positive voltage.
\\

We have shown that STM-LE is an extremely valuable tool for fluorescence measurements of charged species with molecular and even sub-molecular scale precision. In our experiments, we could control the lifetime of the charged molecule and the radiative decay probabilities of the molecule by adjusting respectively the thickness of the insulator separating the molecule from the surface and the plasmonic response of the tip-sample cavity. The ability to address the interplay between the redox and excited states of organic structures with atomic-scale spatial accuracy will be highly valuable to characterize internal charge transfer and exciton separation at covalently linked donor-acceptor molecular dimers and low-band gap copolymers. Eventually, no less than three different luminescence mechanisms have been evidenced for this single system, illustrating the complexity of STM-LE measurements on single molecules.\\

\noindent
\section*{Acknowledgments}
The authors thank Virginie Speisser, Michelangelo Romeo, Jean-Georges Faullumel and Olivier Cregut for technical support. \textbf{Funding:} The Agence National de la Recherche (project SMALL'LED No. ANR-14-CE26-0016-01), the Labex NIE (Contract No. ANR-11-LABX-0058\_NIE), and the International Center for Frontier Research in Chemistry (FRC) are acknowledged for financial support. \textbf{Author Contributions:} All authors contributed jointly to all aspects of this work. \textbf{Competing interests:} The authors declare no competing interests. \textbf{Data and materials availability:} All data needed to evaluate the conclusions in the paper are present in this paper and the supplementary materials.
\\

\end{document}